# One-step formation of plasmonic Cu nanodomains in p-type $Cu_2O$ matrix films for enhanced photoconversion of n-ZnO/p-$Cu_2O$ heterojunctions

*Yerila Rodríguez-Martínez,* [†,‡] *Lídice Vaillant-Roca,* [†] *Jaafar Ghanbaja,* [‡] *Sylvie Migot,* [‡] *Yann Battie,* [§] *Sidi Ould Saad Hamady,* [∥] *David Horwat* [*,‡].

[†] University of Havana, Photovoltaic Research Laboratory, Institute of Materials Science and Technology – Physics Faculty, San Lázaro y L, 10 400 Havana, Cuba.

[‡] Université de Lorraine, CNRS, IJL, F-54000 Nancy, France

[§] Université de Lorraine, LCP-A2MC, Institut Jean Barriol, 1 Blvd. Arago, 57070 Metz, France.

[∥] Université de Lorraine, CentraleSupélec, LMOPS, F-57000 Metz, France.

ABSTRACT. Plasmonic Cu nanoparticles were in-situ grown into a $Cu_2O$ semiconductor matrix by using reactive magnetron sputtering and adjusting the amount of oxygen available during the synthesis in order to prevent the oxidation of part of copper atoms landed on the film surface. Varying only the oxygen flowrate (OFR) and using a single Cu target it was possible to observe the evolution in the simultaneous formation of metallic Cu and $Cu_2O$ phases for oxygen-poor conditions. Such formation is accompanied by the development of the surface plasmon band (SPB) corresponding to Cu, as evidenced by UV-Vis spectrophotometry and spectroscopic ellipsometry. The bandgap values of the elaborated composites containing embedded Cu plasmonic nanodomains were lower than the bandgap of single-phased $Cu_2O$ films, likely due to the higher defect density associated to the nanocrystalline nature of films, promoted by the presence of metallic Cu. The resistivity of the thin films increased with more oxidative deposition conditions and was associated to an increase in $Cu_2O$/Cu ratio and smaller and more isolated Cu particles, as evidenced by high resolution transmission electron microscopy and X-ray diffraction. Photoconversion devices based on the studied nanocomposites were characterized by I-V and spectral photocurrent measurements, showing an increase in the photocurrent density under light illumination as consequence of the plasmonic particles excitation leading to hot carrier's injection in the nearby ZnO and $Cu_2O$ semiconductors.

INTRODUCTION. Metal nanoparticles (NP) have been extensively studied due to their promising optical properties.[1] One of the most interesting feature of metal NP is their ability to exhibit the so-called Localized Surface Plasmon Resonance (LSPR), which refers to a coherent oscillation of the conduction band electrons in nanoobjects interacting with an external electro-magnetic field of the same frequency as the metal plasmon frequency.[2-4] Plasmon modes and energies strongly depend on metal NP parameters like size, shape, inter-particle distance and the refractive index of the surrounding medium.[5, 6]

While metal NP hosting LSPR can be produced by various chemical or physical methods such as electrodeposition,[7] sputtering,[8] laser ablation and electron beam lithography,[6] isolated metal NP, among which Au, Ag, and Cu are known examples,[9] tend to aggregate and some of them may be prone to oxidation. An interesting strategy to avoid these phenomena is to embed them into a proper matrix to better exploit their unique optical behavior and, at the same time, protect them from the surrounding environment. Interfacing semiconductor metal oxides, like ZnO, $Cu_2O$ and $TiO_2$, with metal nanoparticles is of great interest for the enhancement of light harvesting in photocatalysis and photovoltaics applications.[8, 10-13] Nevertheless, most of the studies on such systems involve expensive plasmonic metals (like Au and Ag). Cu is an interesting/inexpensive alternative to consider due to its abundance in the earth crust, nontoxicity and recyclability. Copper being already used in the microelectronic industry for interconnects, copper-based devices would be easy to implement. Furthermore, it has been demonstrated the effective transfer of "hot" charge carriers from this metal to various metal oxide semiconductors like ZnO,[14] TiO$_2$ [15] and Cu$_2$O[16] during illumination, as consequence of localized surface plasmon resonance deexcitation. During the last decade, the number of studies using Cu$_2$O in various applications, e.g., photocatalysis,[17-20] gas sensors [21-23] and solar cells,[17, 24-26] have increased due to its low cost, possibility of including it in large scale production as a result of the abundance of copper and oxygen, its absorption in the visible range of the electromagnetic spectrum and p-type conductivity. Moreover, this semiconductor can be obtained by multiple vacuum and non-vacuum methods such as electrochemical deposition, chemical oxidation, pulsed laser deposition, chemical vapor deposition, sputtering, among others.[17]

Coupling metallic Cu NP to Cu$_2$O and protecting the NP from oxidation by the environment is challenging because it requires that Cu$^0$ and Cu$^+$ oxidation states coexist but are well separated spatially and, simultaneously, that Cu$^+$ plays a protecting role for Cu$^0$. Despite this can be achieved in nanoparticles forming core-shell heterostructures using physical or chemical approaches[7, 27-31], the synthesis of composite films using a process compatible with the microelectronic industry would open up new possibilities such as ease of implementation in future devices.

In this study, we report on a single step route to synthesize nanocomposite films consisting of Cu NP embedded in a Cu$_2$O matrix and exhibiting the LSPR effect. This study tests the hypothesis that by carefully controlling the amount of oxygen available during the condensation of copper vapor to form a thin film, it is possible to restrict the oxidation to a fraction of the copper atoms landed on the growing film surface and to induce the segregation of remaining metallic copper atoms in order to form the sought architecture. For this, we resort to reactive sputtering deposition from a copper target in an argon-oxygen atmosphere. The influence of the elaboration parameters on the microstructure, optical and electrical properties of the synthesized films is investigated. Furthermore, by I-V and spectral photocurrent measurements of $Cu_2O$/Cu nanocomposite-based devices, an increase in the photocurrent is observed due to the hot carrier's injection, compared to devices where plasmonic particles were not included.

RESULTS AND DISCUSSION. *Synthesis and characterization of $Cu_2O$/Cu composites.* XRD measurements allowed to analyze the different phases obtained when the OFR is increased. A mixture of Cu and $Cu_2O$ was found for samples synthesized using 2 and 4 sccm $O_2$, labelled as samples 2 and 4, respectively. In contrast, the X-ray diffractograms of samples synthesized using 5 and 6 sccm $O_2$, respectively labelled as samples 5 and 6, indicate the presence of $Cu_2O$ as a single phase (Figure 1a). A mixture of $Cu_2O$, $Cu_4O_3$ and CuO was obtained for higher OFR (8, 10 and 12 sccm), as confirmed by X-rays diffraction presented in the Supporting Information (Figure S1). In the following we restrain our study to samples 2, 4 and 5, of higher interest for our purpose.

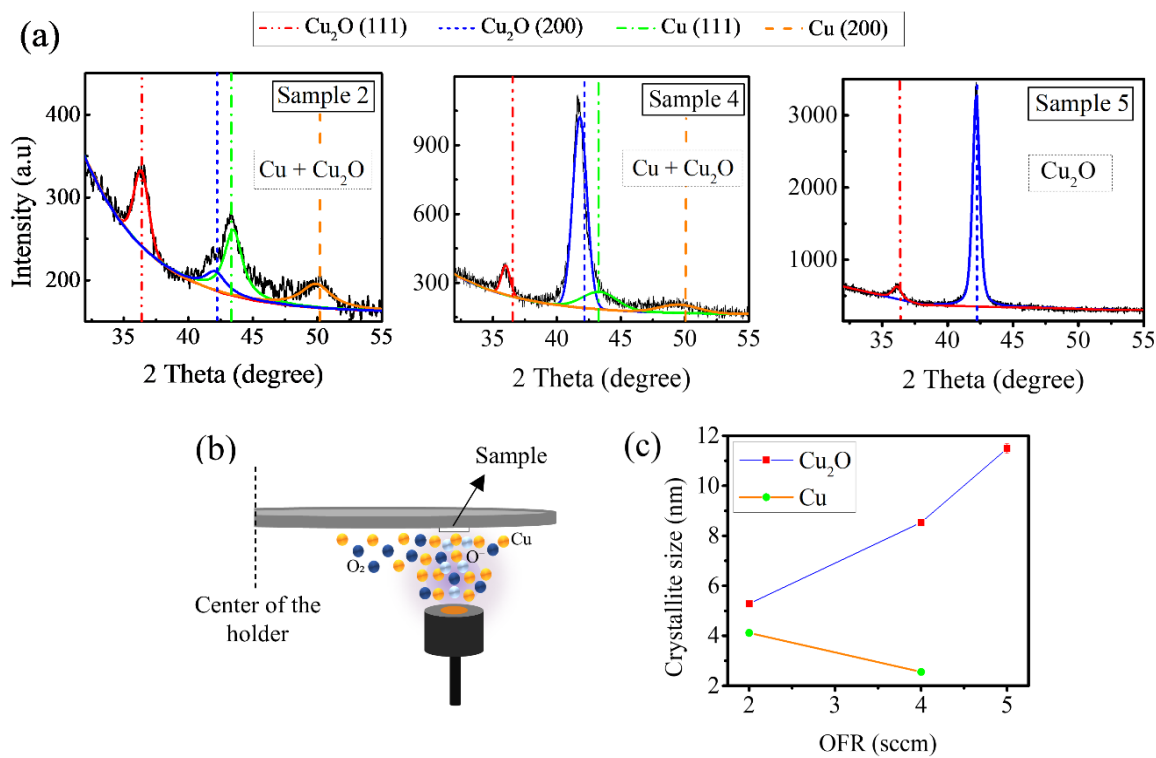

Figure 1. X-ray diffractograms of samples grown at 2 sccm, 4 sccm and 5 sccm of $O_2$ (samples 2, 4 and 5 respectively) with the corresponding deconvolution (Voigt function) of the Cu and $Cu_2O$ peaks

(a). Schematic representation of the incidence of energetic oxygen ions (light blue) which is increased when the OFR increases, leading to higher energy transfer for (200) planes formation (b). Evolution of the average size of crystallites of the (111) and (200) oriented domains of Cu and $Cu_2O$, with the OFR. The colors of the two curves represent those the XRD peaks used for the calculation (c).

The two cubic phases, Cu and $Cu_2O$, in sample 2, coexist with a [111] preferential orientation (JCPDS 00-004-0836 and JCPDS 04-007-9767, respectively) although weaker (200) peaks are also observed. Increasing the OFR to 4 sccm, is accompanied by a noticeable change in the preferential orientation of the $Cu_2O$ phase: the (200) peak becomes much more intense. This behavior becomes even more remarkable as the OFR is further increased to 5 sccm, since only the signal of the $Cu_2O$ phase is observed. A possible explanation for this behavior may be pointed to the surface energy of the (111) and (200) crystallographic planes : the surface energy of the (200) planes is higher than that of (111).[32] In another study related to the manipulation of growth orientation of reactively sputter-deposited $Cu_2O$ films, Wang et al. have obtained the (111) or (200) orientation at high and low sputtering pressure, respectively.[32] As increasing the sputtering pressure lowers the energy of adatoms, the authors proposed that the growth of (100) requires more energetic adatoms. An increasing research interest related to reactive sputter deposition of oxides is the influence of film bombardment by energetic oxygen ions on the film structure. There are indications in the literature that sputtered oxygen atoms can capture a secondary electron to form $O^-$ ions that are accelerated to gain a kinetic energy equivalent to the difference between the plasma potential and the target voltage,[33, 34] i.e., several hundreds of eV. Within this mechanism, the ion-metal flux ratio ($J_{O^-}/J_{Cu}$) incident to the sample increases with the OFR and more energetic oxygen ions impact on the surface (Figure 1b), transferring more energy to the growing film and ease the formation of (200) planes.[35]

In addition to the influence of the OFR on the crystalline orientation, it was possible to notice an evolution in the length of coherent domains of both $Cu_2O$ and Cu phases. Figure 1c shows the evolution with OFR of the average size of the coherent domains along [111] and [100] directions of Cu and $Cu_2O$, respectively, calculated by fitting to Voigt functions and applying the Scherrer's formula. It can be seen an increase in the $Cu_2O$ crystallite size when the OFR increases, while the Cu domains become smaller due to the incremented oxidation.

From the above discussion it can be considered that the oxygen concentration in the samples influence mainly the Cu oxidation states, and therefore, the fractions and nature of phases obtained for different OFR. The energetic oxygen ions, on the other hand, have a more important role in the lattice orientation. This is also supported by the evolution of the preferential orientation dependency with the position of the sample respect to the target axis. More oriented layers were grown close to the target axis, where a higher impinging of oxygen atoms occurs (see Supporting Information, Figure S2).

The incorporation of Cu nanoparticles also has an impact on the optical properties of the grown layers, as evidenced in the absorption spectra extracted from transmittance and reflectance measurements. In Figure 2a a distinctive band centered around 2 eV is observed in samples 2 and 4, while it is absent in the spectrum of sample 5. According to the literature,[36, 37] this band corresponds to the surface plasmon resonance band (SPB) of metallic copper nanoparticles. The position and height of the band, as well as the bandwidth, are related to the shape, size, size distribution, surface state, surface coverage, and surrounding environment of the given nanoparticles.[6] Moreover, due to the interparticle coupling, the proximity of these nanostructures can induce red-shifts and broadening of the SPB.[5] In sample 2, the SPB is broader and red-shifted, compared to sample 4. Then, it's expected to have in that case a higher proximity of the Cu domains with a larger size. The latter was already confirmed by XRD (Figure 1). Additionally, the presence of only one SPB in both cases suggests the presence of rounded structures.[9] The absence of any band that could imply the presence of a plasmonic effect in sample 5 is in agreement with the XRD results, where only $Cu_2O$ phase was observed for this condition. For higher OFR, where there is no presence of Cu, there is neither any absorption peak close to the reported plasmon band of Cu (see Supporting Information, Figure S3).

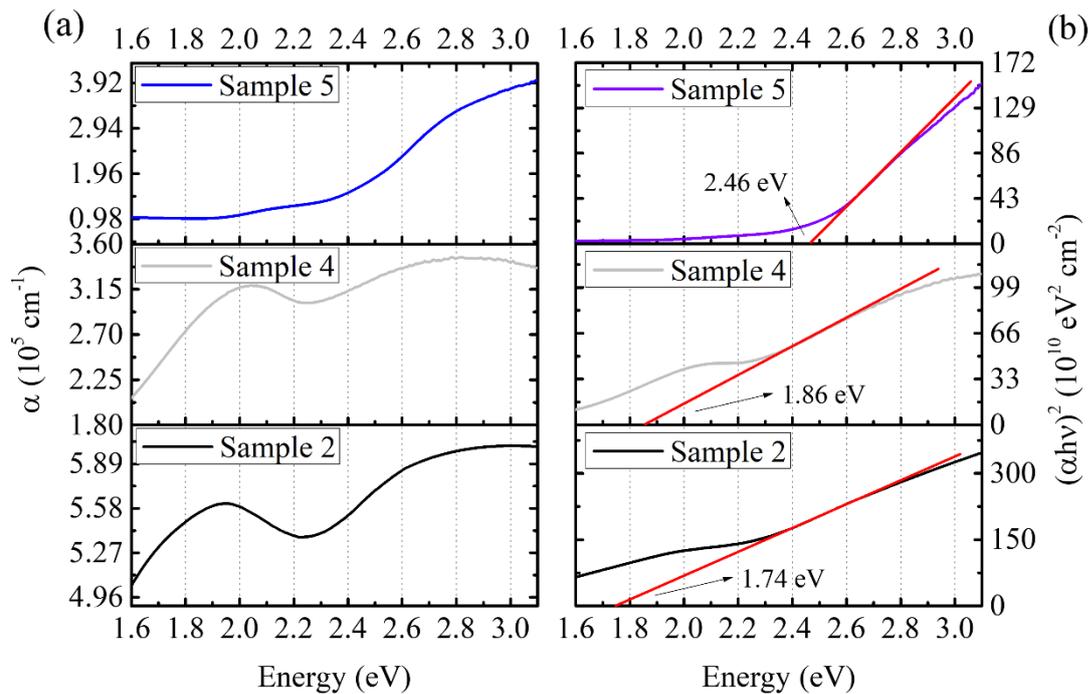

Figure 2. Absorption curves (a) and respective Tauc plots (b) for samples 2, 4 and 5. SPB are observed for samples 2 (around 1.9 eV) and sample 4 (close to 2 eV). For sample 5, no SPB is observed. The bandgap increases as the OFR is higher.

The bandgap values were estimated using the Tauc plot method, assuming that the direct transition takes place in these films (Figure 2b). Sample 5 shows a bandgap value around 2.46 eV, close to the value reported for $Cu_2O$ material.[38, 39] In Sample 2 and sample 4 the values obtained (1.74 eV and 1.86 eV, respectively) cannot be considered as representing $Cu_2O$ matrix bandgap since the presence of the nanoparticles induces absorption related to interband transitions in metallic copper. However, the higher the polycrystalline character of the layer (as in the case of sample 2), the higher the density of grain boundaries, which increases the disorder in the crystalline structure.[35] This disorder is expected to increase the density of localized states close to the CB and VB, and therefore decrease the bandgap value.[40, 41]

Spectroscopic ellipsometry was used to obtain more information about the Cu nanoparticles. The models used for the fitting of the data (Tauc-Lorentz for sample 5 and a sum of two Lorentzian oscillators and a Tauc-Lorentz model for samples 2 and 4) are shown in Figure 3a (Sample 5) and

Figure 3b (Samples 2 and 4) (data in section 4 of the Supporting Information). The imaginary parts of the effective dielectric functions for sample 2 and 4 (Figure 3 (c, d)) show plasmonic modes at 1.48 eV (Sample 2), 1.26 eV (Sample 4) and 1.86 eV (Sample 4). In all cases, a wide peak is observed above 2 eV, which is associated to the band and interband transitions in the $Cu_2O$ matrix and Cu NPs, respectively. Real part of the dielectric function in sample 2 has negative values below 2 eV that, together with the presence of the plasmonic peak in the imaginary part of the dielectric function, suggests a hybrid behavior: metallic (due to the Cu nanoparticles in contact or forming chains) and Cu nanoparticles with plasmonic effect. Moreover, even though this is beyond the scope of the present study, it is worth noting that this sample exhibits epsilon near zero features in the 0.8 to 2.1 eV range that could be of value for various applications. [42-44]

In the case of sample 4, the real part of the function is positive, indicating the LSPR of the particles embedded in dielectric $Cu_2O$.

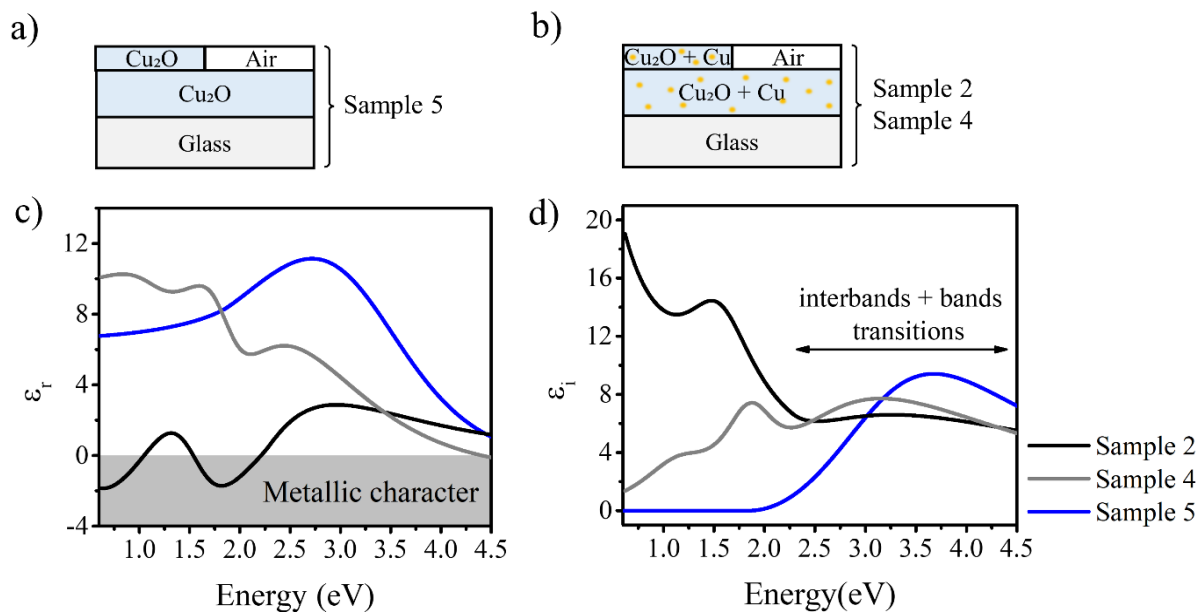

Figure 3. Physical models used for fitting ellipsometric measurements: Tauc-Lorentz (a) and a sum of two Lorentzian oscillators and a Tauc-Lorentz model (b). Real (c) and imaginary (d) parts of the effective dielectric function for samples 2, 4 and 5.

The resistivity measurements of samples 2, 4 and 5 are shown in Figure 4. Low resistivities closer to the value reported for bulk copper, around $10^{-6}$ $\Omega$.cm,[45] correspond to the sample where the presence of this metal is higher (lower OFR: 2 sccm). Then, a progressive increase in resistivity is observed as the formation of $Cu_2O$ increases, until reaching the resistivity reported for this material, in the range from 10 to $10^5$ $\Omega.cm$.[46] Considering this, it is reasonable to think that the main contribution to the resistivity is the formation of more $Cu_2O$ or less Cu phase. Analyzing the two lower OFR, we can consider other factors as well, like more isolated copper particles unable to share conduction electrons through the material and smaller grain size in sample 4 that causes an increase in the scattering by boundaries. The average size of coherence domains estimated from XRD indicates that when the OFR increases to 4 sccm, the length of $Cu_2O$ domains becomes larger but that of Cu domains decreases. In addition, following the assumption that the Cu nanoparticles are more separated in sample 4, it is expected that the resistivity becomes higher for this OFR.

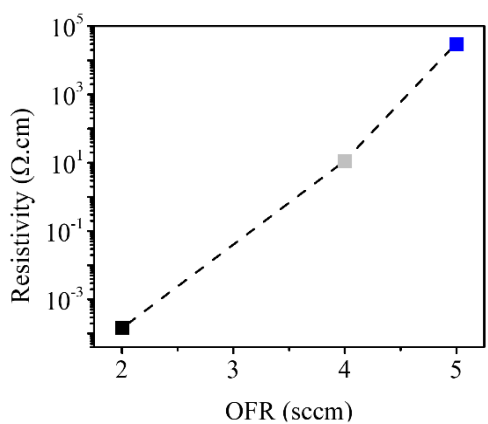

Figure 4. Resistivity evolution with the OFR.

The interpretation of these results is supported by the TEM analysis of samples 2 and 4 shown in Figure 5. The bright and dark field micrographs (Figure 5 (a, b, f, g)) and selected area electron diffraction patterns (SAED) (Figure 5 (d, i)) show two main differences between sample 4 and sample 2: in sample 4 the crystalline orientation is less random and a columnar growth microstructure is observed, in agreement with the XRD measurements. HRTEM micrographs (Figure 5 (c, h)) reveal a decrease in the size of Cu grains when the OFR increases, $(7.29 \pm 2.92)$ nm in sample 2 and $(4.10 \pm 0.68)$ nm in sample 4. This is again consistent with the behavior observed in the domain size calculations (Figure 1).

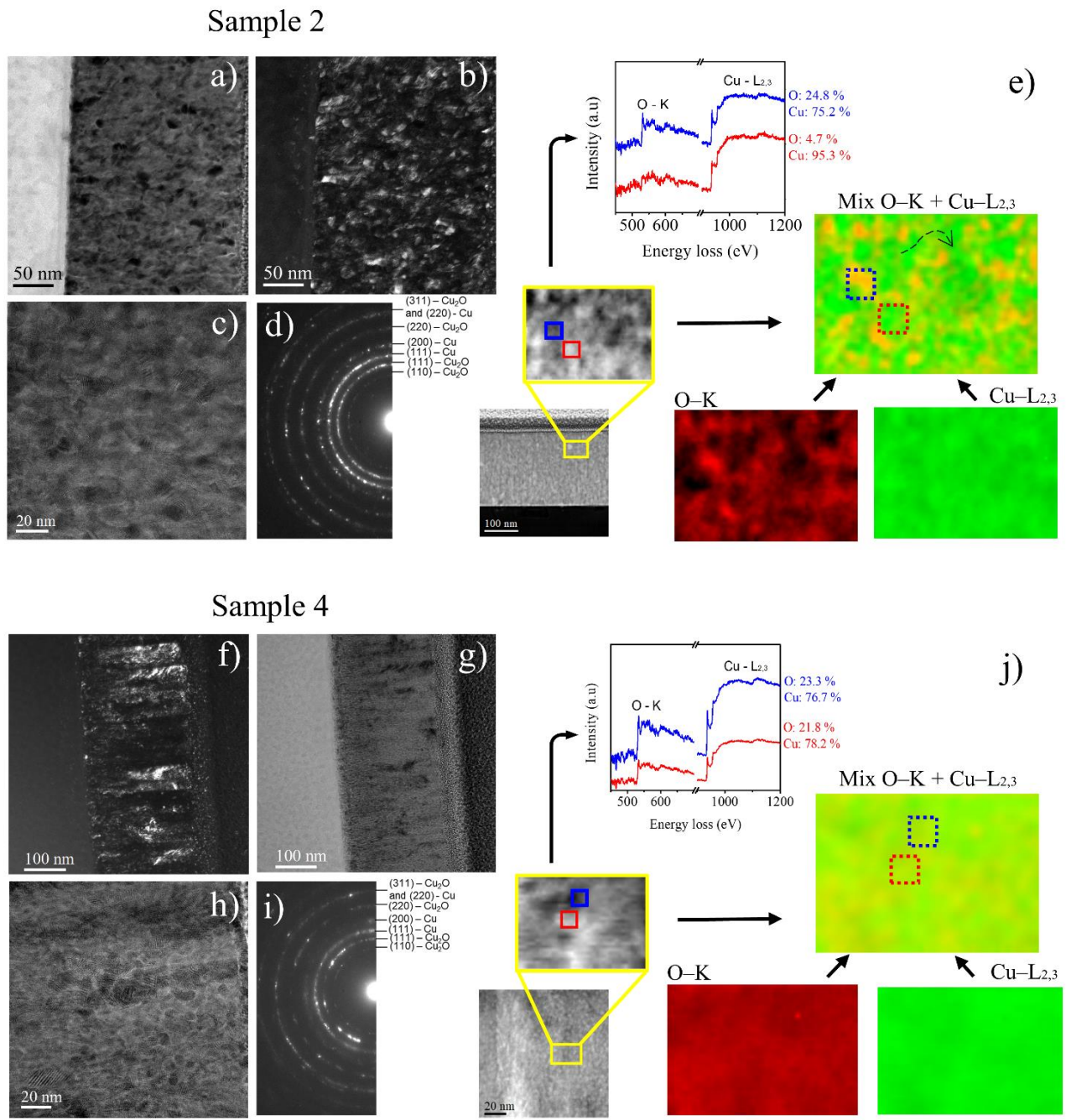

Figure 5. TEM analysis of samples 2 and 4: (a, b, f, g) bright and dark field micrographs where it can be seen the remarked nanocrystalline nature of the sample 2 while a columnar growth is observed for sample 4; (d, i) selected area electron diffraction patterns; (c, h) HRTEM images showing the size distribution of particles; (e, j) high angle annular dark field (HAADF)-STEM micrograph and energy loss near edge structure (ELNES) spectra at O-K and Cu-$L_{2,3}$ edges corresponding to the blue and red square areas in a selected zone of the HAADF micrograph (yellow rectangle). Composition maps of copper (green), oxygen (red) and mixture of both (orange). Dashed arrow in the mixed elemental map of sample 2 (e) show the connection between copper domains.

Figure 5 (e, j) shows the High Angle Annular Dark Field (HAADF)-STEM micrographs taken for the EELS analysis of samples 2 and 4 and the composition maps of Cu, O and the Cu-O mixture, corresponding to the yellow enclosed region. In the analyses of these maps, it is needed to consider the fact that information is obtained from the FIB lamella thickness that exceeds by far the size of crystallites. Hence a superposition of signals is observed. This could affect mostly the study of the Cu phase separately from the $Cu_2O$. However, it was possible to extract useful information. In the mixed composition map of sample 2, the bright green area of the map represents a high copper concentration, while orange zones indicate a higher compensation between Cu and O. This can be seen in ELNES measurements at O-K and Cu-$L_{2,3}$ edges in two different regions (red and blue squares located in the green and orange regions, respectively). In the green region (red spectrum in the ELNES graph) a very weak oxygen edge is observed around 531.9 eV and the copper $L_3$ at 937.9 eV and $L_2$ at 957.9 eV edges are rather flat, indicative of the presence of metallic copper.[47, 48] The orange area (blue spectrum), instead, shows a more defined oxygen edge peak at 532.4 eV as well as a more pronounced copper $L_3$ edge at 937.4 eV, which differentiate the oxidation state $Cu^{+1}$ from $Cu^0$ according to the literature.[49] A closer look at the map, tends to confirm the connection of copper domains into the copper oxide matrix in sample 2 (black dashed arrow). In sample 4, the ELNES spectra are similar and in line with the presence of $Cu^{+1}$ in the two areas, and the different regions are not as easy to distinguish as in the previous case. However, it is possible to notice some green regions corresponding to the lower oxygen concentration, which suggest more isolated particles in sample 4, as suspected above when analyzing the resistivity and optical data. Based on the present study, it is interesting to note that the proposed elaboration process is versatile. Effectively, it enables, by varying a single parameter (the oxygen content in the gas phase), to elaborate pure $Cu_2O$ films, $Cu/Cu_2O$ nanocomposite films with embedded Cu plasmonic nanoparticles or $Cu/Cu_2O$ nanocomposite films with embedded Cu plasmonic nanoparticles and percolating conductive Cu paths.

*Photocurrent enhancement with plasmonic particles incorporation.* Three configurations of devices were fabricated in order to evaluate the impact of incorporating the studied composites films with plasmonic NP (Figure 6). The device Dev_6sccm is a reference device with a p-n junction between ZnO and $Cu_2O$ (grown at 6 sccm) layers. On the other hand, Dev_4sccm and Dev_4sccm/6sccm are two new configurations where the plasmonic-semiconductor composite grown at 4 sccm is replacing the $Cu_2O$ layer and incorporated as a thin layer close to the junction, respectively.

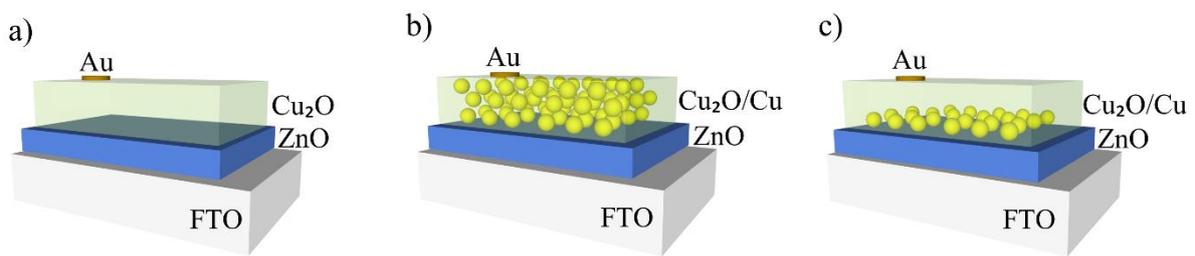

Figure 6: Schemes of the three fabricated devices with and without plasmonic composites (grown at 4 sccm): Dev_6sscm (a), Dev_4sscm (b), Dev_4sscm/6sscm (c).

Current density – voltage (J-V) measurements under dark and illuminated conditions (1 Sun) are shown in Figure 7 (a, b, c). In all the cases, a rectifying behavior typical of a p – n junction is observed. Upon illumination, all of them showed an increase in current density at 0 V, indicating the occurrence of a typical carrier collection process, which is higher for devices in which plasmonic nanodomains were incorporated. Although this was an expected result, it is interesting to notice that by decreasing the thickness of the layer composed by plasmonic nanodomains (Dev_4sccm/6sccm), responsible for the injection of extra carriers, the $J_{sc}$ is almost 4.5 times higher than for the reference device and 2 times higher than for Dev_4sccm (Figure 7d).

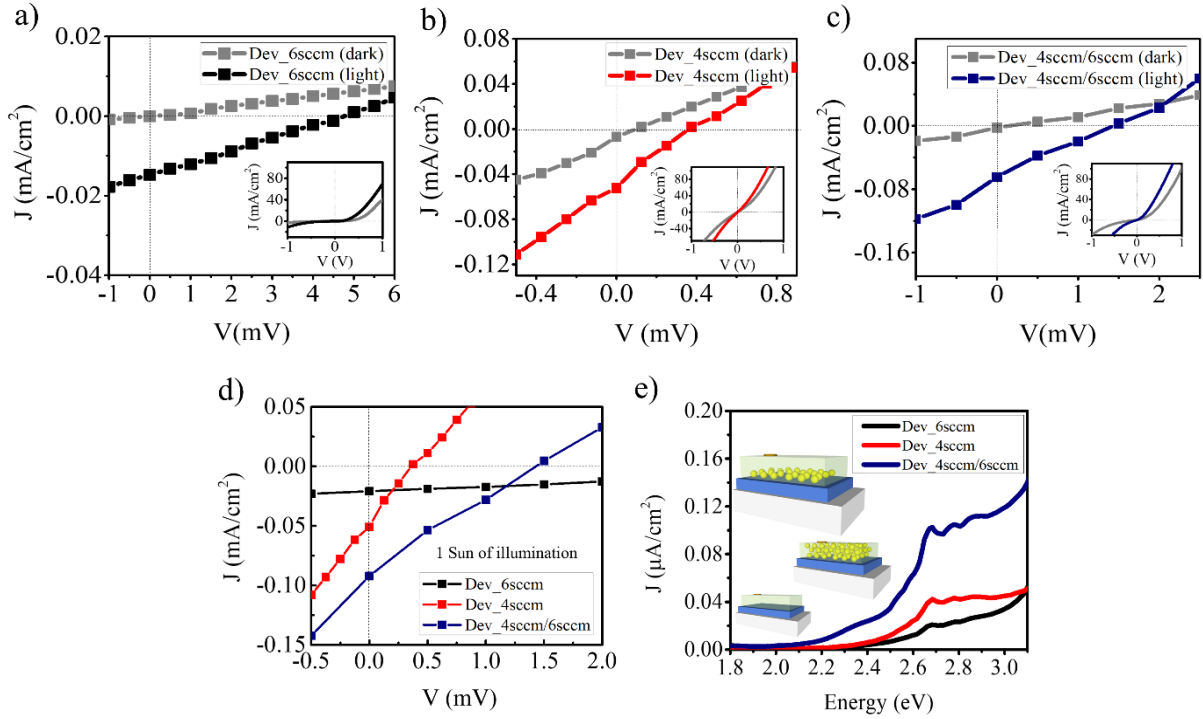

Figure 7. Current density vs voltage measurements in dark and at 1 sun of illumination (a-d). Modulated photocurrent spectroscopy (e).

$J_{sc}$ is proportional to the minority carrier diffusion length and the absorption coefficient of the material.[50] In nanostructured films with many grain boundaries, like the studied plasmonic composites, a relatively high defect density is expected, which could decrease the carrier diffusion length. Then, although the generation rate should be increased in Dev_4sccm due to a higher absorption coefficient and more carrier's injection induced by the plasmonic effect, low values of carrier diffusion length avoid significant increase in the current density. That explains why Dev_4sccm/6sccm, an intermediate architecture between the other two configurations, is associated to an improvement in the charge generation and limited decrease of the diffusion length, allowing to collect more photogenerated carriers than the reference device. This configuration could also be favored by the strong local field enhancement [51] around the metal nanoparticles, allowing photocarriers being generated and separated close to the collection junction area [3, 52] (Figure 8 (a-c)).

The above interpretation can be supported by analyzing the values of $V_{oc}$, which inversely depends on the saturation current $J_0$: [53]

$$V_{oc} = \frac{kT}{q} \ln\left(\frac{J_{sc}}{J_0} + 1\right) \qquad (1)$$

where $k$ is the Boltzmann constant, $T$ the temperature and $q$ the electron charge. The reverse saturation density current $J_0$ gives a measure of the recombination in the device related to the diffusion lengths. If $J_0$ increases more quickly than $J_{sc}$, then a lower value of $V_{oc}$ is obtained.

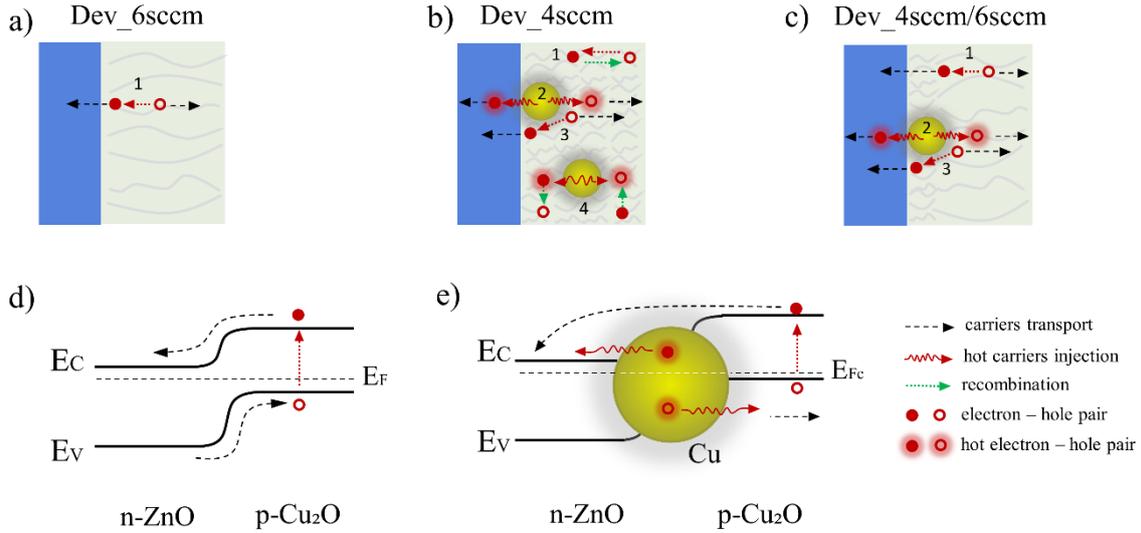

Figure 8. Schematic of the transfer, injection and recombination of carriers in different devices (a-c) and corresponding band alignments (d-e) under illumination, where Ec and Ev are the conduction and valence bands, respectively, and $E_F$ and $E_{Fc}$ are the Fermi level of the system without and with nanoparticles (composite), respectively. Processes related to the photocurrent generation (a-c): electron – hole pair generation and recombination (1); hot carrier's injection from plasmonic NP (2); carrier separation by strong local field enhancement around the metal NP (3); annihilation of injected hot carriers by recombination (4). Band coupling between n – ZnO, p – $Cu_2O$ and Cu NP.

Modulated photocurrent measurements support the fact that plasmonic particles induce a positive effect on the generation of hot carriers (Figure 7e), particularly in Dev_4sccm/6sccm. The photocurrent response in this case, shows a broad band between 2 eV and 2.5 eV approximately, which is not seen in the other two configurations. This band is interpreted as the result of the injection of hot carriers into the semiconductors after photons with energies greater than 2 eV excite the metal

NP. Effectively, according to calculations done by Tagliabue et. al. [54], for energies above 2 eV, highly-energetic hot holes far below the Cu Fermi level (yellow dashed line in Figure 8e) are produced with high probability of being injected in the p-type material, while hot electrons with much lower energy are located just above the Cu Fermi level and can be injected to the n-type material. As represented in the Figure 8e, the Cu Fermi level is aligned with the Fermi levels of $Cu_2O$ and ZnO. This alignment allows the excited hot electron and hot hole to reach the conduction band of ZnO and the valence band of $Cu_2O$, respectively. This is possible because both the hot electron and the hot hole have enough energy to overcome the energy barrier with the corresponding semiconductor. The position of the Fermi level of ZnO is the same in Figure 8d (Dev_6sccm) and Figure 8e (Dev_4sccm/6sccm) while the Fermi level of $Cu_2O$ in Figure 8e (Dev_4sccm/6sccm) can shift towards the valence band due to a higher presence of acceptor defects like copper vacancies.

The collection of injected hot holes into the $Cu_2O$ matrix in Dev_4sccm could be limited by the low mean free path of hot holes [54] and a high density of recombination centers. In Dev_4sccm/6sccm, hot holes, after being transferred to the thin $Cu_2O$ matrix, are immediately transported to the interfaced $Cu_2O$ layer of lower defect density, and subsequently collected (Figure 8 (b, c)). On the other hand, it is worth mentioning that from these measurements, it seems that the bandgap of the $Cu_2O$ matrix merely changes between the devices, staying around 2.4 eV. This supports the hypothesis that the values obtained in the Tauc plot for samples 2 and 4 are affected by the interbands transitions occurring in the Cu nanoparticles.

The above study highlights the possibility to prepare in a one-step method hybrid plasmonic NP/p-$Cu_2O$ layers triggering photoconversion enhancement when inserted in n-ZnO/p-$Cu_2O$ heterojunctions. Furthermore, the photocurrent and responsivity values (see Supporting Information, Figure S5) of the devices studied are comparable or even higher than recently studied photodetectors and solar devices where metal NPs are included (Table 1). It is worth to notice that many of the configurations reported include nanostructured scaffolds like nanorods (NRs) / nanowires (NWs), which allow to increase the effective area for the charge photogeneration and separation and may positively impact in the photoconversion. On the other hand, it can be seen that the use of Au/Ag expensive metals in these designs is predominant. As stated above, Cu based configurations likely correspond to more interesting choice from the perspective of raw material abundance and cost. Many other studies have included plasmonic nano-objects in photocatalysis devices [55, 56] and sensors [57] showing the promising potential of all-oxide heterostructures with plasmonic NPs like presented in this work.

Table 1. Figures of merit of ZnO, $Cu_2O$ and plasmonic NPs-based photodetectors and solar devices.

| Heterostructure | Photocurrent | Responsivity | Detection Range | Ref. |
|---|---|---|---|---|
| Au NPs/$Cu_2O$ [1] | ~1.32 µA/cm² (0 V) | ~22 µA/W (0V) | 532 nm (VIS) | 58 |
| Al:ZnO/Ag Nps/$Cu_2O$ [2] | 21.41 mA/cm² (0 V) | -- | VIS | 59 |
| ZnO NRs/Au NPs [2] | 15.8 µA/cm² (0 V) <br> 74 µA/cm² (0 V, 450°C) | -- | VIS | 60 |
| ZnO/$CdMoO_4$/Au NPs [3] | ~289.1 nA (5 V) | 321 mA/W (5 V) | 350 nm (UV) | 61 |
| ZnO/Cu NPs [4] | 950 mA/cm² (-6 V) | 100 mA/W (-6 V) | 400-800 nm (VIS) | 14 |
| ZnO NRs/Au NPs/$Al_2O_3$ [5] | ~160 nA (0 V) | 6.8 mA/W (0 V) | 365 nm (UV) | 62 |
| ZnO NWs/Au NPs [6] | -- | 510 mA/W (0 V) | 940 nm (NIR) | 63 |
| ZnO NRs/Al NPs [7] | ~1 µA (5 V) | 0.4 mA/W (5 V) | 325 nm (UV) | 64 |
| ZnO NRs/Cu NPs [7] | ~0.1 mA (5 V) | 46 mA/W (5 V) | 325 nm (UV) | 64 |
| ZnO/Cu NPs-$Cu_2O$ [2] | 52.4 µA/cm² (0 V) <br> 188.9 mA/cm² (1 V) | 0.45 mA/W (0 V) <br> 539.6 mA/W (1 V) | VIS | this work |
| ZnO/Cu NPs-$Cu_2O$/ $Cu_2O$ [2] | 64.5 µA/cm² (0 V) <br> 151.2 mA/cm² (1 V) | 0.62 mA/W (0 V) <br> 531.3 mA/W (1 V) | VIS | this work |

Power densities of the light sources: [1] 60 mW/cm², [2] 100 mW/cm², [3] 0.15 mW/cm², [4] 7 W/cm², [5] 0.6 W/cm², [6] 0.35 mW/cm², [7] 2.17 W/cm².

*Performance enhancement of the nanocomposites.*

Some considerations about possibilities to improve the performance of the composites grown using an OFR of 4 sccm worth to be discussed. The low crystallinity of the matrix in nanoparticle-incorporated composites grown by sputtering has also been seen in previous works, and is associated to local strain and impurities interfering with the crystal growth.[65] Post-annealing could decrease the density of such defects and improve the matrix crystallinity, favoring the enhancement of the diffusion length. Yet, as seen in other studies, this would likely be accompanied by a segregation of metal to the surface and agglomeration of nanoparticles.[66] A careful control of the annealing atmosphere would also need to be considered because annealing in oxidative atmosphere would lead to uncontrolled decay in the density of acceptor states in the matrix and to particle oxidation. Another way to try improving the performance could be through fine tuning of the oxygen concentration during synthesis. Especially, according to the spectral photocurrent results, the proximity of the SPB to the absorption band of $Cu_2O$, plays an important role in the enhancement of the photocurrent. This proximity depends on particle size, itself affected by the OFR. Within the tested conditions, the optimal OFR is likely between 4 sccm and 5 sccm, but the difference in the position of the SPB should not be noticeable using the same experimental conditions. Increasing the flux of sputtered metallic atoms, through an increase of the discharge current applied to the copper target, could widen the working window for OFRs leading to transition from percolated to isolated copper NPs in $Cu_2O$. Therefore, it could give more freedom to precisely adjust the size of particles.

CONCLUSIONS. In this work, we proposed a single-step route to synthesize plasmonic copper nanodomains embedded in a p-type semiconducting $Cu_2O$ matrix that trigger enhanced photoconversion when inserted at the interface of n-ZnO/p-$Cu_2O$ heterojunctions. The enhancement finds its origin in hot carrier's injection and collection following deexcitation of localized surface plasmon resonance (LSPR) at plasmonic Cu nanodomains. The single-step route proposed is based on the control of oxygen partial pressure or oxygen flow rate during reactive sputtering of copper, a method that can be easily scaled up and is compatible with the microelectronic industry.

EXPERIMENTAL SECTION. *Materials and methods.* Cu-O films were grown on glass substrates (1 cm x 2.5 cm) by reactive DC magnetron sputtering of a Cu target (50 mm diameter and 3 mm thick with a purity of 99.99 %, Kurt J. Lesker Company). After mechanical cleaning with ethanol, the substrates were loaded inside a 40-l sputtering chamber and placed on a circular substrate-holder, at 7.5 cm from the substrate-holder rotation axis, facing the Cu target axis. The substrate-holder was put in rotation with a frequency of 3.5 Hz to obtain a good thickness homogeneity of the films. The distance from the target to the holder was 5 cm for all experiments. Prior to deposition the chamber was evacuated using a primary mechanical and a secondary turbomolecular pumps.

For deposition of the Cu-O films, the argon flowrate was fixed at 50 sccm (standards cubic centimeters per minute) and the flowrate of the reactive $O_2$ gas (OFR), was varied from 2 sccm to 12 sccm. The base pressure in the chamber was around 0.04 Pa and the total pressure during deposition was between 0.3 to 0.5 Pa, applying a current of 0.15 A. Surface activation was done by radiofrequency polarization (1 min) before each growth process to enable good adhesion of the deposited film to the glass substrate. The deposition was accomplished close to room temperature, no intentional heating was applied. According to the corresponding flowrate, samples were named: sample 2 (2 sccm), sample 4 (4 sccm), sample 5 (5 sccm), sample 6 (6 sccm), sample 8 (8 sccm), sample 10 (10 sccm) and sample 12 (12 sccm).

Three configurations of devices were fabricated in order to evaluate the impact of plasmonic Cu NP in the current generation of optoelectronic devices. The first one was a reference cell consisting of a ZnO (~100 nm) / $Cu_2O$ (~200 nm) p-n junction (labeled Dev_6sccm). In a second device the $Cu_2O$ layer was replaced by a composite of $Cu_2O$ and plasmonic copper NP obtained at 4 sccm of oxygen flowrate (~200 nm) (labeled Dev_4sccm). The last configuration includes a very thin interlayer of composite (~10 nm) between the ZnO n-type material and the $Cu_2O$ p-type material (labeled Dev_4sccm/6sccm). In all the cases, the ZnO layer was deposited on a previously cleaned FTO substrate by Reactive Magnetron Sputtering, using a Zn target (99.995 % of purity, Kurt J.

Lesker Company) and 8 sccm of OFR. The distance of the target to the holder plate was 5 cm and the sample was placed at 3 cm from the center of the holder axis to achieve the better conductivity.[47] The argon flow rate was fixed at 50 sccm and the applied current was 0.07 A, reaching a working pressure of 0.4 Pa. Same procedure described in previous paragraph was applied to deposit the Cu-O layers. In the case of the last configuration, a two-step route was followed: a very thin layer (10 nm) was first deposited using 4 sccm of OFR and then the OFR was increased to 6 sccm until the layer reached 200 nm of thickness. Finally, an Au back contact was deposited by magnetron sputtering, using an Au target (99.99 % of purity, Neyco Company), located at 5 cm from the holder plate, where the sample was located in the target axis. The current applied was 0.06 A and the working pressure was 0.5 Pa.

*Characterization.* An X-ray diffractometer (Bruker D8 Advanced) in Bragg/Brentano configuration (Cu Kα X-ray source, wavelength λ = 0.15406 nm) was employed to study the crystalline structure of the samples. Optical measurements were performed using a Cary 5000 UV−vis-NIR spectrophotometer and the film thickness, used for absorption coefficient and resistivity calculations, was measured by a DektakXT stylus profilometer (vertical resolution of 0.1 nm). Spectroscopic ellipsometry was performed using a UVISEL ellipsometer using 3 incidence angles (50°-60°-70°). A Tauc-Lorenz model was used to extract the dielectric constants of pure $Cu_2O$ while a sum of two Lorentzian oscillators and a Tauc-Lorentz model was used to extract the dielectric constants of $Cu_2O$/Cu composite films. The resistivity of the samples was measured using the four-point probe method in the linear configuration via a Keithley 2700 Multimeter and a Keithley 237 Source Measure Unit. Transmission electron microscopy (TEM) investigations were carried out using a JEM - ARM 200F Cold FEG TEM/STEM operating at 200 kV coupled with a GIF Quantum 965 ER. The samples were prepared using a focused ion beam with a dual-beam SEM-FIB FEI Helios NanoLab 600i. Using TEM mode, bright field and dark field images were collected. With HRTEM, the size of particles was estimated and, using STEM mode, energy loss near edge structure (ELNES) spectra at the O-K and Cu-$L_{2,3}$ edges as well as Cu and O composite composition maps were obtained.

The I-V measurements in dark and under 1 sun of light intensity were performed using a Keithley 2400 Source Meter Unit, a halogen lamp of 150 W and 24 V whose intensity was regulated by a DSC-Electronics Programmable DC Power Supply using a Calibrated Reference silicon solar cell from Radboud University Nijmegen. Spectral photocurrent was measured using modulated photocurrent spectroscopy system equipped with a xenon lamp with a high stability regulated power system (300 W from Lot-Oriel), a monochromator (Omni-λ 300i model with a 1200 g/mm grating) giving a monochromatic light in the 300-1100 nm wavelength range with a resolution of 0.1 nm, a Standford Research current preamplifier (model SR570) and a lock-in amplifier (model SR850).

ASOCIATED CONTENT

**Supporting Information.** Additional characterization data (XRD patterns for flow rates higher than 5 sccm, XRD patterns of samples grown at 2, 4 and 5 sccm varying the position respect to the target axis, absorption coefficient and Tauc plots for flow rates higher than 5 sccm, spectroscopic ellipsometry measurements and fitting, responsivities of Dev_6sccm, Dev_4sccm and Dev_4sccm/6sccm, deposition rate evolution with OFR) (PDF).

AUTHOR INFORMATION

**Corresponding Author**

* Email: david.horwat@univ-lorraine.fr

**ORCID**

Yerila Rodríguez-Martínez: 0000-0001-5314-0101

Lídice Vaillant-Roca: 0000-0003-1552-7449

Jaafar Ghanbaja: 0000-0003-2870-0570

Sylvie Migot: 0000-0003-3803-9931

Yann Battie: 0000-0003-2421-2684

Sidi Ould Saad Hamady: 0000-0002-0480-6381

David Horwat: 0000-0001-7938-7647

**Author Contributions**

The manuscript was written through contributions of all authors. All authors have given approval to the final version of the manuscript. DH [‡], LVR [†] planned and supervised the study. YRM [†, ‡] preformed the experiments, processed the data and wrote the manuscript. Y.B. [§] performed the ellipsometry measurement and modeling. J.G. [‡] performed the TEM analysis and S.M. [‡] prepared the TEM lamella. SOSH [l] carried out the modulated photocurrent spectrometry measurements.

**Funding Sources**

Cooperation and Cultural Action Service, French Embassy in Havana, Cuba.

Functional Thin Films and Applications team, Institute Jean Lamour.

**Notes**

The authors declare no competing financial interest.

ACKNOWLEDGMENT

We thank the support provided by the French Embassy in Havana through its Cooperation and Cultural Action Service (SCAC), Campus France, the Institute Jean Lamour (UMR 7198 CNRS / University of Lorraine), particularly to the team Functional Thin Films and Applications. We also thank to the Erasmus Mundus Master Program AMASE for its support and LMOPS laboratory

ABBREVIATIONS

LSPR, localized surface plasmon resonance; OFR, oxygen flow rate; NP, nanoparticles; SPB, surface plasmon band; XRD, x-ray diffraction; DC, direct current; FTO, fluorine-doped tin oxide; HRTEM, high resolution transmission electron microscopy; TEM, transmission electron microscopy; STEM, scanning transmission electron microscopy; FFT, fast Fourier transform pattern; FIB, focused ion beam; ELNES, energy loss near edge structure; EELS, electron energy loss spectroscopy; CB, conduction band; VB, valence band.

Supporting Information

One-step formation of plasmonic Cu nanodomains in p-type Cu$_2$O matrix films for enhanced photoconversion of n-ZnO/p-Cu$_2$O heterojunctions

*Yerila Rodríguez-Martínez,* [†,‡] *Lídice Vaillant-Roca,* [†] *Jaafar Ghanbaja,* [‡] *Sylvie Migot,* [‡] *Yann Battie,* [§] *Sidi Ould Saad Hamady,* [∥] *David Horwat* [*,‡].

[†] University of Havana, Photovoltaic Research Laboratory, Institute of Materials Science and Technology – Physics Faculty, San Lázaro y L, 10 400 Havana, Cuba.

[‡] Université de Lorraine, CNRS, IJL, F-54000 Nancy, France

[§] Université de Lorraine, LCP-A2MC, Institut Jean Barriol, 1 Blvd. Arago, 57070 Metz, France.

[∥] Université de Lorraine, CentraleSupélec, LMOPS, F-57000 Metz, France.

**Corresponding Author**

* Email: david.horwat@univ-lorraine.fr

1. XRD patterns of Cu-O layers grown at higher OFR (6, 8, 10 and 12 sccm) are shown in Figure S1.

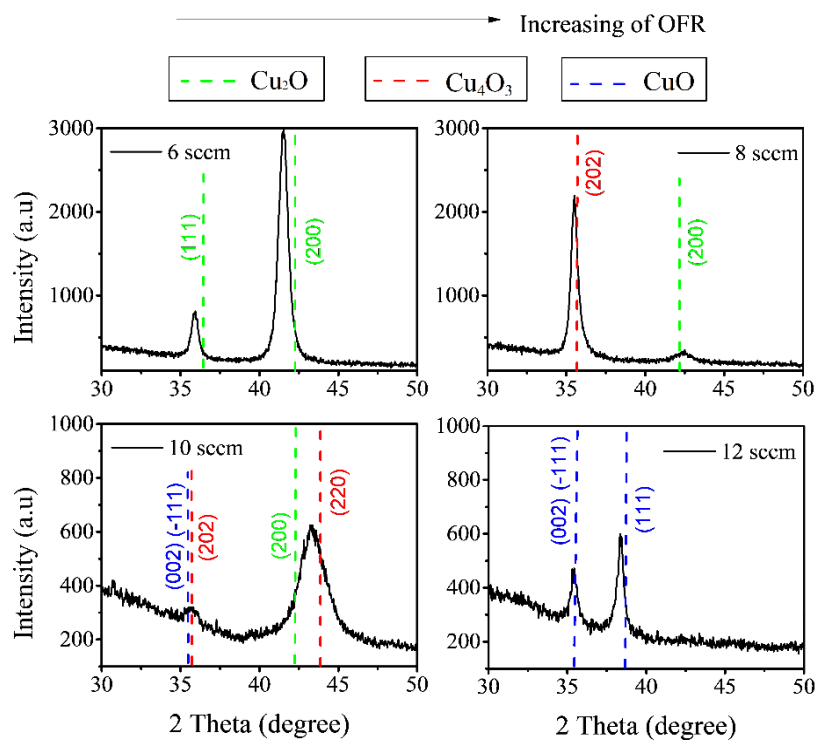

Figure S1. XRD patterns of samples grown at 6 sccm, 8 sccm, 10 sccm and 12 sccm. With the increasing of the OFR different phases of copper oxides appear, since the less oxidation state ($Cu_2O$) to the higher one (CuO). The metastable phase $Cu_4O_3$ can be seen as well. The evolution is in agreement with the reported results in literature.[39]

2. XRD patterns of Cu-O layers grown at 2 sccm, 4 sccm and 5 sccm varying the position of the samples respect to the target axis.

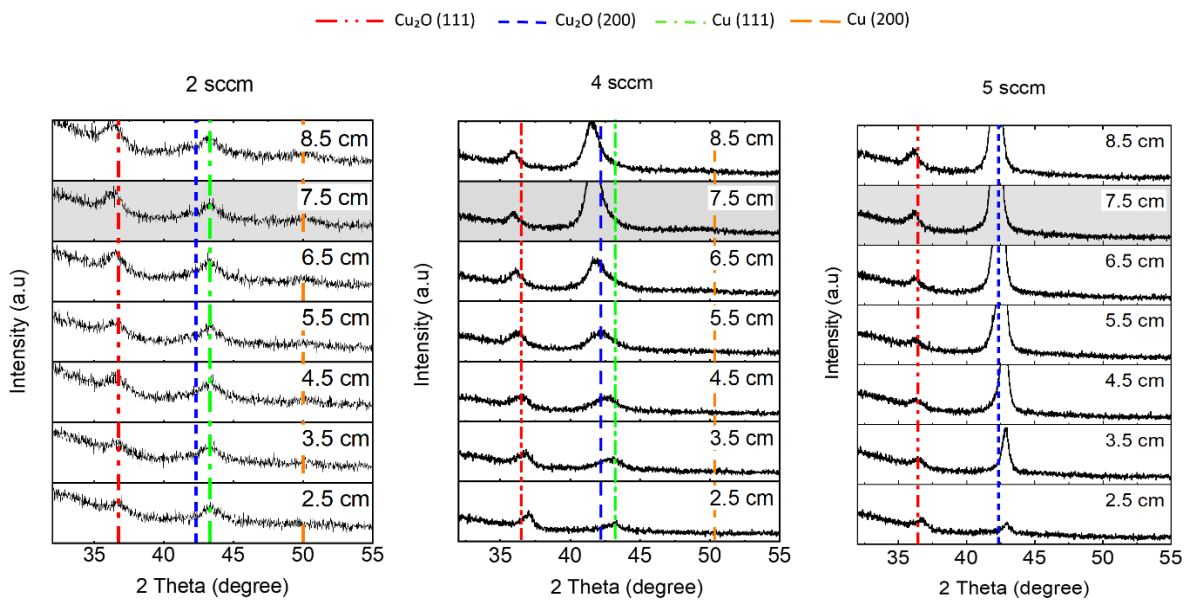

Figure S2: XRD patterns for different OFR (2, 4 and 5 sccm) changing the position of the samples, measured respect to the center of the sample holder. The shadow position (7.5 cm) corresponds to the target axis.

3. In Figure S3 it can be seen the absorption coefficient of Cu-O layers grown at higher OFR (6, 8, 10 and 12 sccm).

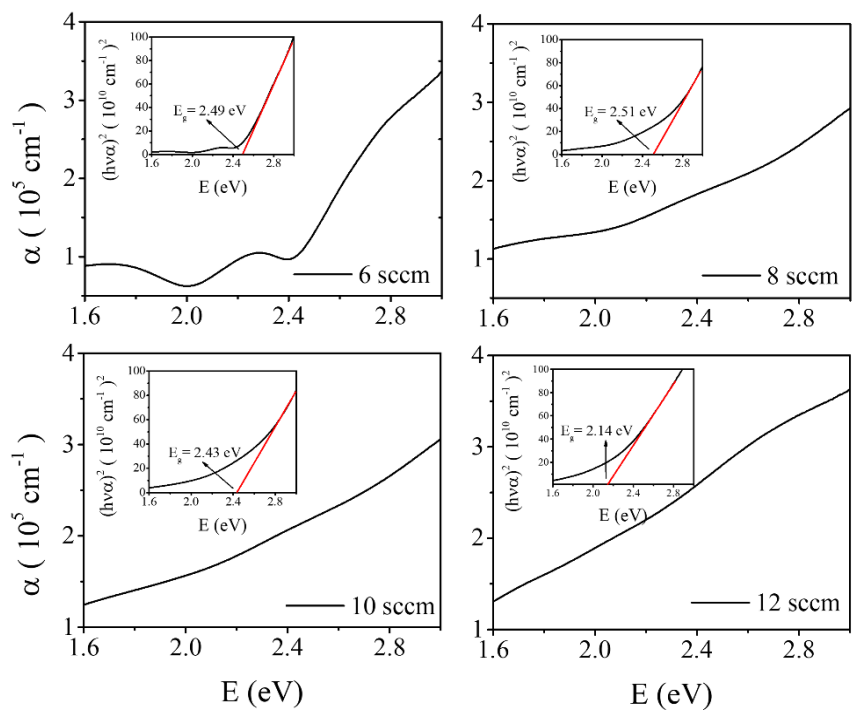

Figure S3. Absorption coefficients for samples 6, 8, 10 and 12. SPBs are not observed. The peaks below 2.4 eV in sample 6 appear as a consequence of the light interference in transmittance measurements.

4. Spectroscopic Ellipsometry

Ellipsometric measurements were made at 3 angles of incidence 50°, 60° and 70°. There were measured 2 parameters, Is and Ic, which depend on the ellipsometry angles. The model used for sample 5 (Tauc Lorentz) is composed of a $Cu_2O$ layer on a glass substrate with a roughness layer (50% layer - 50% air), while in the model used for samples 2 and 4 (a sum of two Lorentzian oscillators and a Tauc-Lorentz model) the $Cu_2O$ layer includes small metal nanoparticles inside, as it was showed in Figure 4 (a, b). The measurements and fit of the data are shown in Figure S4.

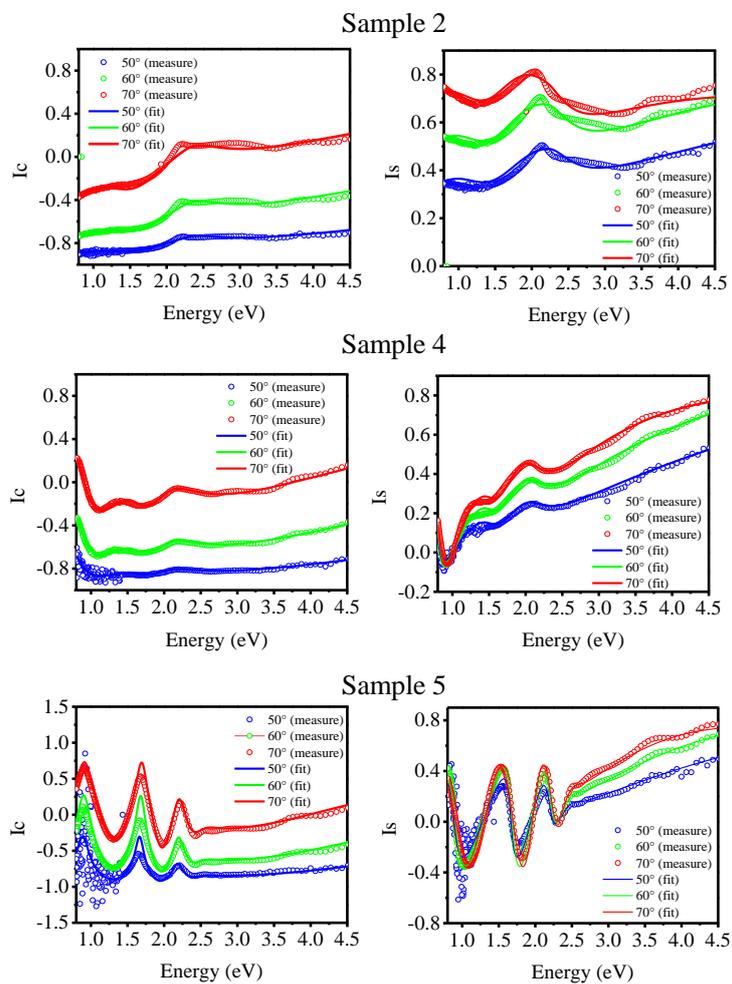

Figure S4. Ellipsometric measurements for the three samples studied and their respective fit using the selected models.

The calculated thickness of the $Cu_2O$ and $Cu_2O$ + Cu layers, as well as the roughness layers are shown in table 1.

Table S1: Calculated thickness of the films and their corresponding roughness layers.

| Sample | $Cu_2O$ and $Cu_2O + Cu$ layers thickness | Roughness layers thickness |
|---|---|---|
| 2 | 199 nm | 5 nm |
| 4 | 245 nm | 1 nm |
| 5 | 275 nm | 6.25 nm |

5. Responsivity of devices

The responsivities of the devices (Figure S4) were calculated using the formula $R = \frac{J}{P}$, where $J$ is the photocurrent density calculated by subtracting the dark current density from the current density under illumination with a power density ($P$) of 100 mW/cm$^2$.

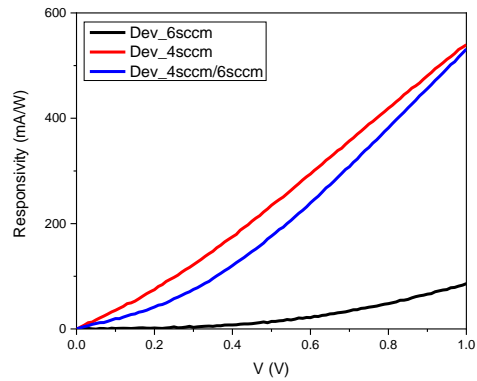

Figure S5: Responsivities calculated from the I-V measurements of the three fabricated devices: Dev_6sccm, Dev_4sccm and Dev_4sccm/6sccm.

6. Deposition rate evolution with the OFR.

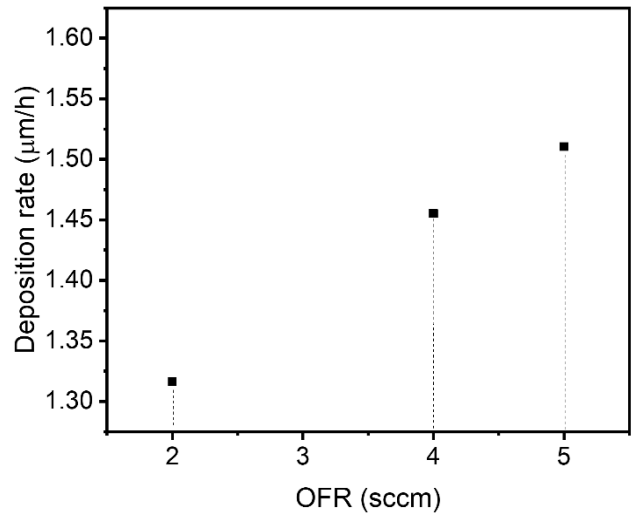

Figure S6: Deposition rate of samples 2, 4 and 5 as function of the OFR.